\documentstyle[aps,prl,multicol,psfig,graphics]{revtex}
\tighten
\begin{document}
\wideabs{
\title{Influence of the pseudogap on the superconductivity-induced 
phonon renormalization in high-T$_c$ superconductors}

\author{S. Varlamov and G. Seibold}
\address{Institut f\"ur Physik, BTU Cottbus, PBox 101344, 
         03013 Cottbus, Germany}
\date{\today}
\maketitle

\begin{abstract}
We investigate the influence of a d-density wave (DDW) gap on the
superconductivity-induced renormalization of phonon frequency
and linewidth. The results are
discussed with respect to Raman and inelastic neutron scattering
experiments. It turns out that the DDW gap can enhance the range of
frequencies for $q=0$ phonon softening depending on the underlying band
structure. Moreover we show that an
anisotropic  'd-wave' pseudogap can also contribute  to the
q-dependent linewidth broadening of the 340cm$^{-1}$ phonon in
YBa$_2$Cu$_3$O$_7$.
\end{abstract}
}

\narrowtext
The origin of the pseudogap is presently a strongly debated issue
in the field of high-T$_c$ superconductivity and has been detected 
by numerous techniques in all the cuprates \cite{TIMUSK}. 
One of the ideas which received considerable attention
involves the existence of preformed pairs \cite{RANDERIA}.
This idea finds support in continuous evolution 
of angle-resolved photoemission spectra
(ARPES)  from the normal to the superconducting (SC) state \cite{ARPES}. 
However, another appealing scenario suggests that
the pseudogap could (not only) arise from pairing in the
particle-particle channel, but also from different scattering
mechanisms (like CDW fluctuations) which become singular
near optimal doping.
Indeed there is increasing experimental evidence \cite{TALLON,BOEB}
that the peculiar properties
of the cuprates, both in the normal and the
SC phase are related to the occurrence 
of a Quantum Critical Point (QCP) located near (slightly above)
the optimal charge carrier concentration. Besides singular scattering
at the QCP a further direct consequence of this scenario is the prediction 
of an ordered state in the underdoped
regime. Various types of orderings have been proposed each of them breaking 
different kinds of symmetries. For example, Castellani et al. \cite{CAST} have 
proposed incommensurate charge-density waves (ICDW) which break translational
invariance. Varma \cite{VARMA} considers circulating currents which
(besides time-reversal symmetry) break a rotational invariance of the 
copper oxygen lattice but preserve translational invariance.
In one of the latest proposals Chakravarty et al. \cite{CHAKRA} have 
investigated d-density wave (DDW) order.
This state is made of staggered currents
which break parity and time-reversal symmetry as well as translational 
invariance by one lattice constant and rotation by $\pi/2$.
All of these proposals have in common that the corresponding
order parameter induces a gap in the single-particle
spectrum which has the same anisotropy as observed in ARPES
experiments.  
For the circulating current state this has been shown 
in Ref. \cite{VARMA2} and for the ICDW phase in Ref. \cite{GOETZ}.
In case of the DDW state the pseudogap anisotropy is explicitly put in
the quasiparticle hamiltonian.

In this paper we consider the influence of an anisotropic single-particle
gap on the frequency shift and linewidth of phonons in the SC state. 
Strong phonon self-energy effects in high-T$_c$ cuprates have been
analyzed first by Zeyher and Zwicknagel \cite{ZEYHER} within Eliashberg theory.
For an isotropic s-wave superconductor they
obtained softening for phonons with frequencies 
$\omega < 2 \Delta$ and hardening for 
phonons with $\omega > 2 \Delta$. In addition the linewidth of the phonon 
mode is only affected for $\omega > 2\Delta$ where it is additionally broadened
due to the opening of the SC (s-wave) gap.    
The Zeyher-Zwicknagel approach has been generalized to anisotropic
gap functions and dispersions relevant for the cuprates in Refs. 
\cite{CARBOTTE,BILL}. The main difference with respect to the s-wave
model is the appearance of an additional line broadening for mode
frequencies $\omega < 2\Delta_0$ since in a d-wave superconductor the 
density of states (DOS) remains finite for energies within the maximum 
SC gap. 

Our investigations are based on the DDW state since from the formal
point of view this scenario offers a simple way of introducing
an anisotropic gap into the problem. However, we want to stress that
most of the results presented below should also (at least qualitatively) 
hold for other QCP proposals as long as the corresponding
pseudogap leads to an analogous quasiparticle dispersion. 
In this context it is interesting to note that a DDW-type
gap has also been considered for the ICDW scenario \cite{BENFATTO}. 
Starting point is the following hamiltonian
\begin{eqnarray}
H&=&\sum_{{\bf k}\sigma} [\epsilon({\bf k})-\mu] 
c_{{\bf k}\sigma}^{\dagger}c_{{\bf k}\sigma} + i 
\sum_{{\bf k}\sigma} \chi({\bf k}) c_{{\bf k+Q}\sigma}^{\dagger}
c_{{\bf k}\sigma} \nonumber \\
&+& \sum_{{\bf k}} \Delta({\bf k}) \lbrack 
c_{{\bf k}\uparrow}^{\dagger}c_{{\bf -k}\downarrow}^{\dagger}+
c_{{\bf -k}\downarrow}c_{{\bf k}\uparrow}\rbrack \label{H0}
\end{eqnarray}
where $\chi({\bf k})=\chi_0[\cos(k_x)-\cos(k_y)]/2$
and $\Delta({\bf k})=\Delta_0[\cos(k_x)-\cos(k_y)]/2$
denote the DDW and SC gap respectively and ${\bf Q}=(\pi,\pi)$.  
The dispersion $\epsilon_{\bf k}$ will be specified later.
In the basis  
$\Psi_k^{\dagger}= (c_{{\bf k}\uparrow}^{\dagger},c_{{\bf -k}\downarrow},
c_{{\bf k+Q}\uparrow}^{\dagger},c_{{\bf -k-Q}\downarrow})$ 
the hamiltonian eq. (\ref{H0}) is given by
\begin{displaymath}
   {H}_{\bf kk}=\left( \begin{array}{cccc} 
(\tilde{\varepsilon_{k}}-\tilde{\mu_k}) & \Delta_{k} & i\chi_{k} & 0 \\
\Delta_{k} & -(\tilde{\varepsilon_{k}}-\tilde{\mu_k}) & 0 & i\chi_{k} \\
-i\chi_{k} & 0 & -(\tilde{\varepsilon_{k}}+\tilde{\mu_k}) & -\Delta_{k} \\
0 & -i\chi_{k} & -\Delta_{k}  & (\tilde{\varepsilon_{k}}+\tilde{\mu_k})\\
    \end{array} \right).
\end{displaymath}
where
$\tilde{\varepsilon_k}=(\varepsilon_k-\varepsilon_{k+Q})/2$ and
$\tilde{\mu_k}=\mu-(\varepsilon_k+\varepsilon_{k+Q})/2$.
Diagonalization yields the four eigenvalues 

\begin{figure}
\begin{center}
\hspace{3cm}{{\psfig{figure=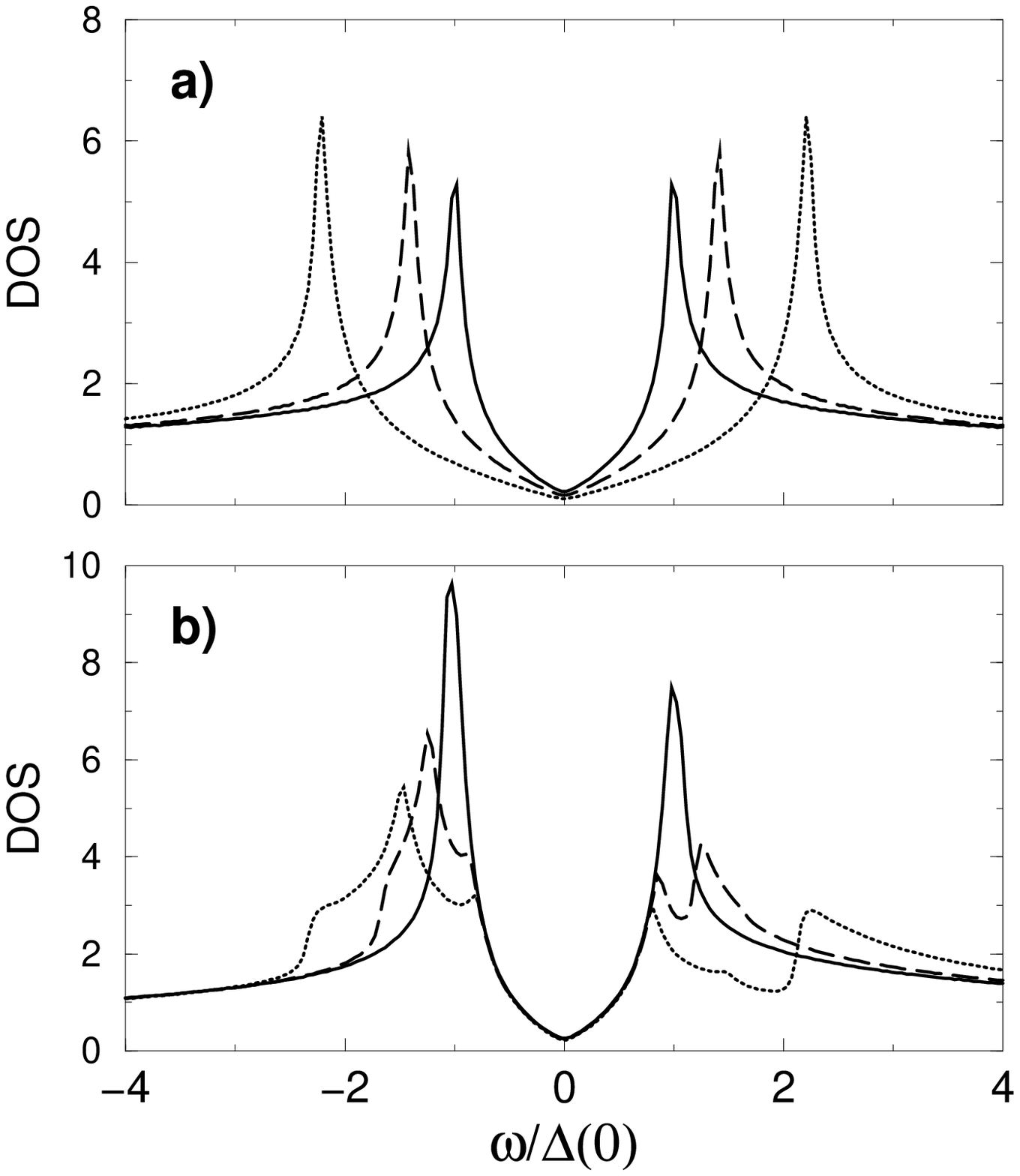,width=6.5cm}}}
\end{center}
\vspace{-0.2cm}
{\small FIG. 1. Density of states in the superconducting state 
$(\Delta_0=23meV)$
for three values of pseudogap: 
$\chi_0/\Delta_0=0$ - solid curve,$\chi_0/\Delta_0=1$ - dashed curve,
$\chi_0/\Delta_0=2$ - dotted curve. 
a) Nearest-neighbor tight-binding model at half filling;   
b) Energy dispersion from Ref. \protect\cite{NORMAN}.}
\end{figure}

$\pm E_{1,2}(k)=\pm\sqrt{(\tilde{E_k} \pm \mid \tilde{\mu_k} \mid )^2
+\Delta_{k}^2}$,
where $\tilde{E_k}=\sqrt{\tilde{\varepsilon_k}^2+\chi_k^2}$.
Note that the wave vector is now within the reduced zone of the 
DDW state \cite{CHAKRA}.

The superconductivity induced phonon renormalization is obtained
from the imaginary-time ordered charge-charge correlation function
$\Pi(q,i\omega)=\langle T_{\tau}{\rho_q}(i\omega){\rho_{-q}(-i\omega)}\rangle$
via $\delta \omega=g^2 \Pi({\bf q},\omega+i\delta)$ 
and the correlator is analytically continued
to real frequencies at the end of the calculation.
Throughout the paper we restrict ourselves to a momentum independent
electron-phonon coupling constant g in order to focus on the essential
contribution from the pseudogap anisotropy.

We start our analysis by calculating the renormalization of
$q=0$ phonons. In this case the charge susceptibility reads as

\begin{displaymath}
\Pi(q=0,i\omega)=\frac{1}{N}\sum_{k \atop s=1,2}
\frac{\Delta_k^2[1-2n_s(k)]}{E_s(k)}
\frac{4}{[(i\omega)^2-(2E_s(k))^2]}
\end{displaymath}

where $n_s(k)=(e^{\beta E_s(k)}+1)^{-1}$.
Note that within our approach the q=0 phonon self-energy is only
finite in the SC state for arbitrary value of the pseudogap. 
In order to investigate the
influence of the pseudogap on the phonon renormalization above T$_c$
one should include vertex corrections to $\Pi(q=0,i\omega)$
which however is beyond the scope of the present paper.

Obviously the underlying dispersion plays a
significant role in determining the properties of $\Pi(q=0,\omega)$.   
Fig. 1 shows the density of states (DOS) for two model band dispersions 
which will be considered below. The DOS is evaluated
in the SC state ($\Delta_0=23meV$ for all following results)
and for three values of the pseudogap. 
Fig. 1a corresponds to $\varepsilon_k^a = -2t[\cos(k_x)+cos(k_y)]$ 
with $t=0.5eV$ i.e.
complete nesting for the DDW scattering vector at half-filling.
Due to particle-hole symmetry the chemical potential $\mu$ is not altered 
in this case upon increasing the pseudogap. Thus for $\mu=0$ $\chi(k)$ and 
$\Delta(k)$ work cooperatively and the DOS shows a single
'leading edge' gap at energy $2\sqrt{\Delta_0^2 + \chi_0^2}$. 
Fig. 1b shows the DOS for a tight-binding dispersion 
$\varepsilon_k^b$ recently proposed by
Norman \cite{NORMAN} in order to analyze neutron-scattering and
ARPES data in the SC state of underdoped bilayer cuprates.
In this case the gap edges of SC gap and pseudogap are
separated since the former opens at the chemical potential whereas the
pseudogap is related to the energy where the dispersion is most
susceptible to DDW scattering.
Moreover the DDW state induces a Fermi surface which consists of
four pockets centered around $(\pm \pi/2,\pm \pi/2)$.
With increasing DDW gap the pockets shrink towards the nodes thus
decreasing the effective SC gap in the DOS (cf. Fig. 1b).

For both dispersions Fig.2 depicts the real and imaginary part 
of $\Pi(q=0,\omega)$ as a function of $\omega/(2\Delta_0)$ 
and for zero temperature. 
In case of zero pseudogap the frequency shift 
and linewidth broadening correspond to the results of Refs. 
\cite{CARBOTTE,BILL} i.e. crossover from hardening to softening at frequencies
$\omega \approx 2\Delta_0$. In addition the d-wave gap results in 
a linewidth broading also for $\omega < 2\Delta$ due to the finite
density of states in the SC gap.
However, consider the case of finite DDW gap and the underlying dispersion
$\varepsilon_k^a$ (left panel).
It turns out that the sign 
change in $\delta \omega$ now occurs at frequencies 
$2\sqrt{\Delta_0^2 + \chi_0^2}$

\begin{figure}
\begin{center}
\hspace{6cm}{{\psfig{figure=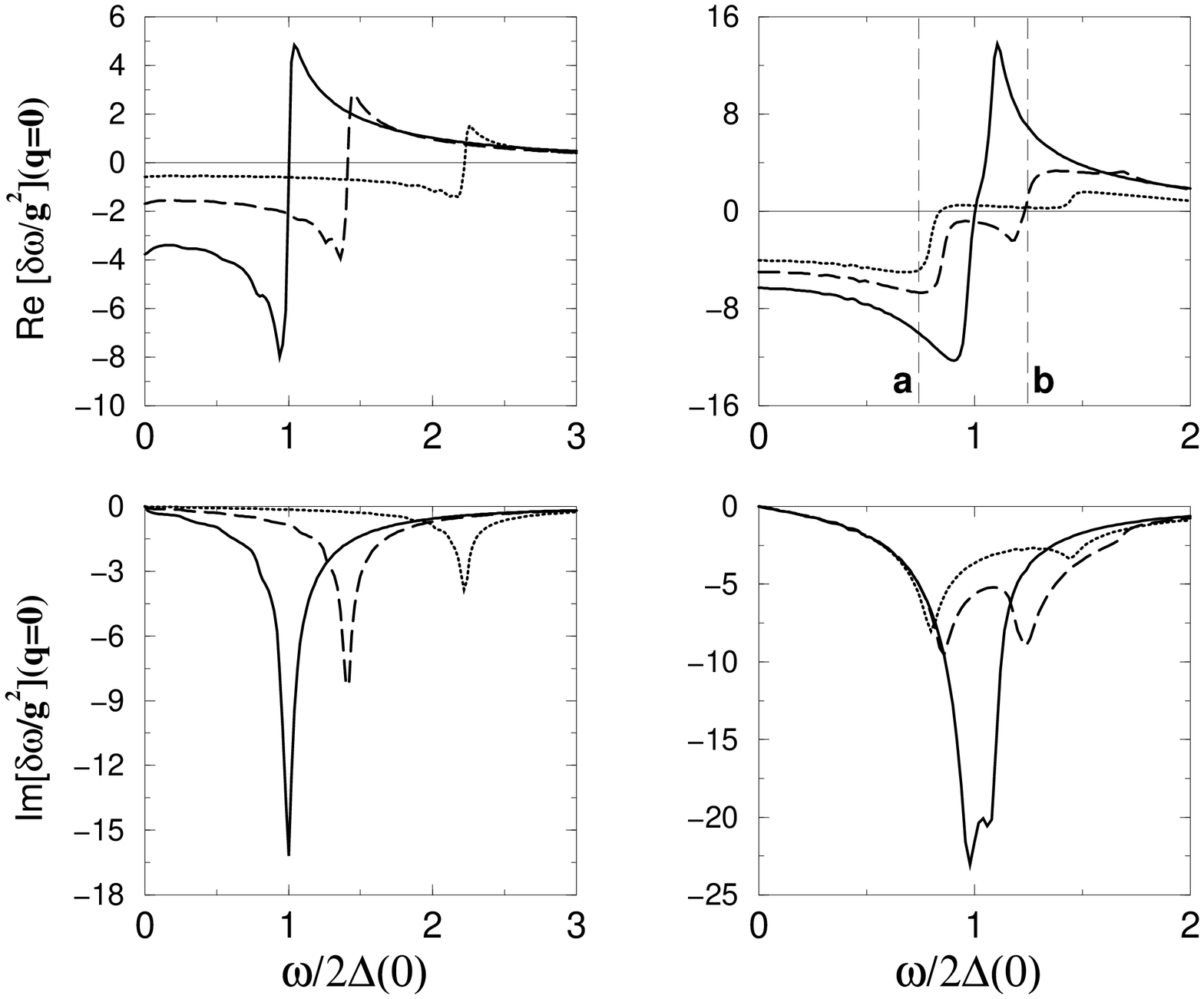,width=7.5cm}}}
\end{center}
\vspace{-0.2cm}
{\small FIG. 2. Real and imaginary part of $\Pi(q=0,\omega)$
corresponding to the DOS shown in Fig. 1a (left panel)
and Fig. 1b (right panel).
Labeling of the curves is the same than in Fig.1. The dashed lines
in the upper right plot indicate the phonon frequencies discussed 
in the text.}
\end{figure}

following the behavior in the DOS.  
Moreover with increasing pseudogap the phonon linewidth
broadening rapidly decreases and correspondingly also the 
maximum phonon frequency shift.
In case of the dispersion $\varepsilon_k^b$ (right panel)
the situation is more complex and can be analyzed best by 
first considering the linewidth broadening which follows closely the
behavior of the DOS. 
Upon increasing the DDW gap  $Im \Pi(q=0,\omega)$
develops a double peak structure \cite{NOTE}, where the low energy
excitation is due to SC and the other one is due to the DDW 'leading
edge' in the DOS. This structure is reflected in  $Re \Pi(q=0,\omega)$
by the two step like features below and above $2\Delta(0)$. It turns out that
at least up to $\chi/\Delta(0)=1$ the DDW gap enlarges the frequency range
where phonon softening occurs. In addition the 'hardening peak' becomes 
strongly
suppressed by a finite value of the pseudogap most notably exemplified in case
of $\chi/\Delta(0)=2$ where $Re \Pi(q=0,\omega)$ is practically zero 
above $\omega \sim 2\Delta(0)$.

We now turn to the analysis of the temperature dependence of the phonon
renormalization which we evaluate for the dispersion $\varepsilon_k^b$. 
Fig. 3 shows the real and imaginary part
of $\Pi(q=0,\omega)$ for the two frequencies
$\omega_{a,b} = 2\Delta(0)\pm \Delta(0)/2$ (indicated by dashed lines in
Fig. 2b).  For simplicity, we adopt the BCS temperature dependence 
for both SC and DDW order parameters, i.e. 
$\Delta_0(T)=\sqrt{1-(T/T_c)^2}$,$T_c=80K$ 
and $\chi_0(T)=\sqrt{1-(T/T^*)^2}$,$T^*=150K$ (note that the temperature
dependence of $\chi_0(T)$ has no significant influence on the result). 
For zero DDW gap the a-phonon initially hardens 
close to T$_c$ where the 'effective' SC gap is well below  $\omega_a$
in agreement with Ref. \cite{BILL}.
This hardening becomes completely suppressed upon increasing the
 DDW gap, however, there is still significant softening for $T\rightarrow 0$.
Correspondingly the peak in $Im \Pi(q=0,\omega)$ is also reduced
by the opening of the pseudogap.
As already discussed above the hardening of the b-phonon is much more 
affected by the DDW gap. In fact the frequency shift is almost completely
suppressed for $\chi/\Delta(0)=2$ where also the strongly reduced DOS in  
the pseudogap reduces the phonon scattering and thus the 
corresponding linewidth.

Our results can be compared with doping dependent Raman scattering
experiments where upon underdoping the increasing pseudogap
should have a significant influence on the phonon renormalization. 
The most pronounced features are found for the 340 cm$^{-1}$ phonon 
in YBa$_2$Cu$_3$O$_{7-x}$ which undergoes a strong softening and 
hardening of the 440cm$^{-1}$ and 500cm$^{-1}$ modes below the SC
transition temperature. 
The particular role of doping for the 
Raman modes below T$_c$ was investigated by Altendorf et al. 
\cite{ALTENDORF} and more recently by Limonov et  al. \cite{LIMONOV}.
It turns out that the superconductivity induced changes in linewidth
and frequency for the 340cm$^{-1}$ B$_{1g}$ mode depend very sensitively
on the charge carrier concentration near optimal doping. 
With increasing underdoping the phonon renormalization becomes rapidly 

\begin{figure}
\begin{center}
\hspace{6cm}{{\psfig{figure=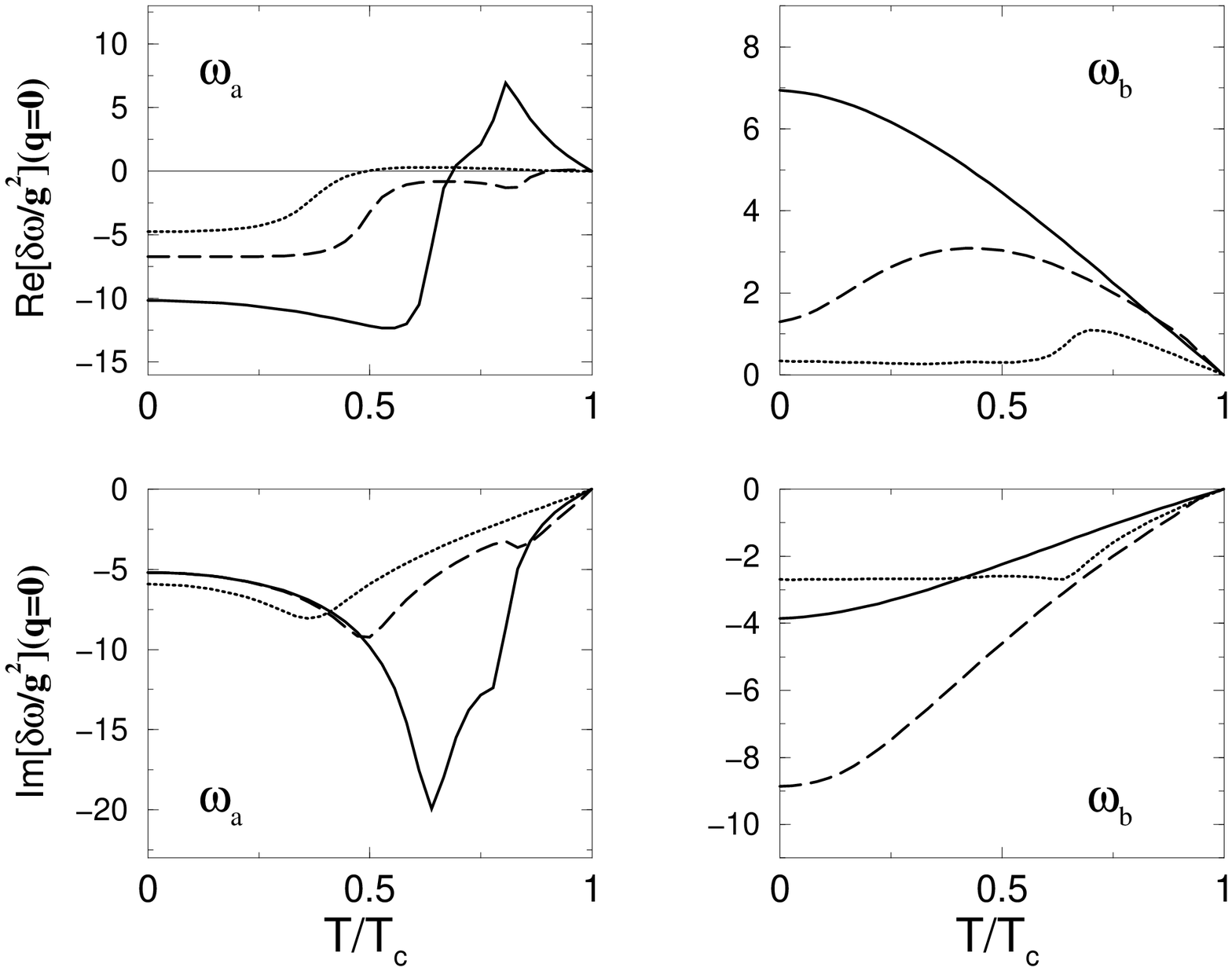,width=7.5cm}}}
\end{center}
\vspace{-0.2cm}
{\small FIG. 3. Temperature dependence of frequency shift and linewidth
broadening for two frequences $\omega_{a,b} = 2\Delta(0)\pm \Delta(0)/2$
(left and right panel). Labeling of the curves is the same than in Fig.1.}
\end{figure}

suppressed in contrast to the overdoped regime where the data of
Ref. \cite{LIMONOV} do not show a pronounced difference of frequency
shift to the optimally doped case. 
Only the linewidth broadening gradually
decreases from the overdoped to the optimally doped sample and vanishes
for the underdoped system. 
Analagous behavior is observed for the 430cm$^{-1}$ and 500cm$^{-1}$
modes which show hardening for the optimal and overdoped samples
while frequency shifts are suppressed in the underdoped regime.
Thus these features can be (at least qualitatively) accounted for by 
evaluating the phonon renormalization with an anisotropic pseudogap 
as described above. 

We now proceed in calculating the influence of the DDW gap on the 
q-dependent phonon renormalization in the SC state
(based on dispersion  $\varepsilon_k^b$).  
For simplicity we restrict ourselves to zero temperature 
where the charge-charge correlation function is given by
\begin{displaymath}
\Pi(q,i\omega)=\frac{1}{2N}\sum_{k \atop s,t=1,2}
\Theta_{s,t} \Omega_{s,t} 
\frac{E_s(k+q)+E_t(k)}{(i\omega)^2-(E_s(k+q)+E_t(k))^2}.
\end{displaymath}

The coherence factors read as

\begin{eqnarray*}
\Theta_{s,t}&=&1+(-1)^{s+t}
\frac{\tilde{\varepsilon_k} \tilde{\varepsilon}_{k+q} +\chi_k\chi_{k+q}}
{\tilde{E_k}\tilde{E}_{k+q}} \,\,\,\,\, \text{and} \nonumber \\
\Omega_{s,t}&=&1-
\frac{[(-1)^s \tilde{E}_k-|\tilde{\mu}_k|]
[(-1)^t \tilde{E}_{k+q}-|\tilde{\mu}_{k+q}|] 
-\Delta_k\Delta_{k+q}}{E_s(k)E_t(k+q)}. 
\end{eqnarray*}

Since we are interested in the change of phonon frequency and linewidth
upon entering the SC state we consider in the following the
difference $\Pi^{\Delta > 0}(q,i\omega)-\Pi^{\Delta=0}(q,i\omega)$.
Fig. 4 depicts the corresponding real and imaginary part 
for $\omega_a = 2\Delta(0)-\Delta/2$ along
the $[1,0]-$ and $[1,1]$ direction for different values of $\chi(0)$.  
Note that the peaklike structures at small q are due to the 
two-dimensionality of the Fermi surface (see e.g. \cite{MARSIGLIO,BILL2})
and 

\begin{figure}
\begin{center}
\hspace{6cm}{{\psfig{figure=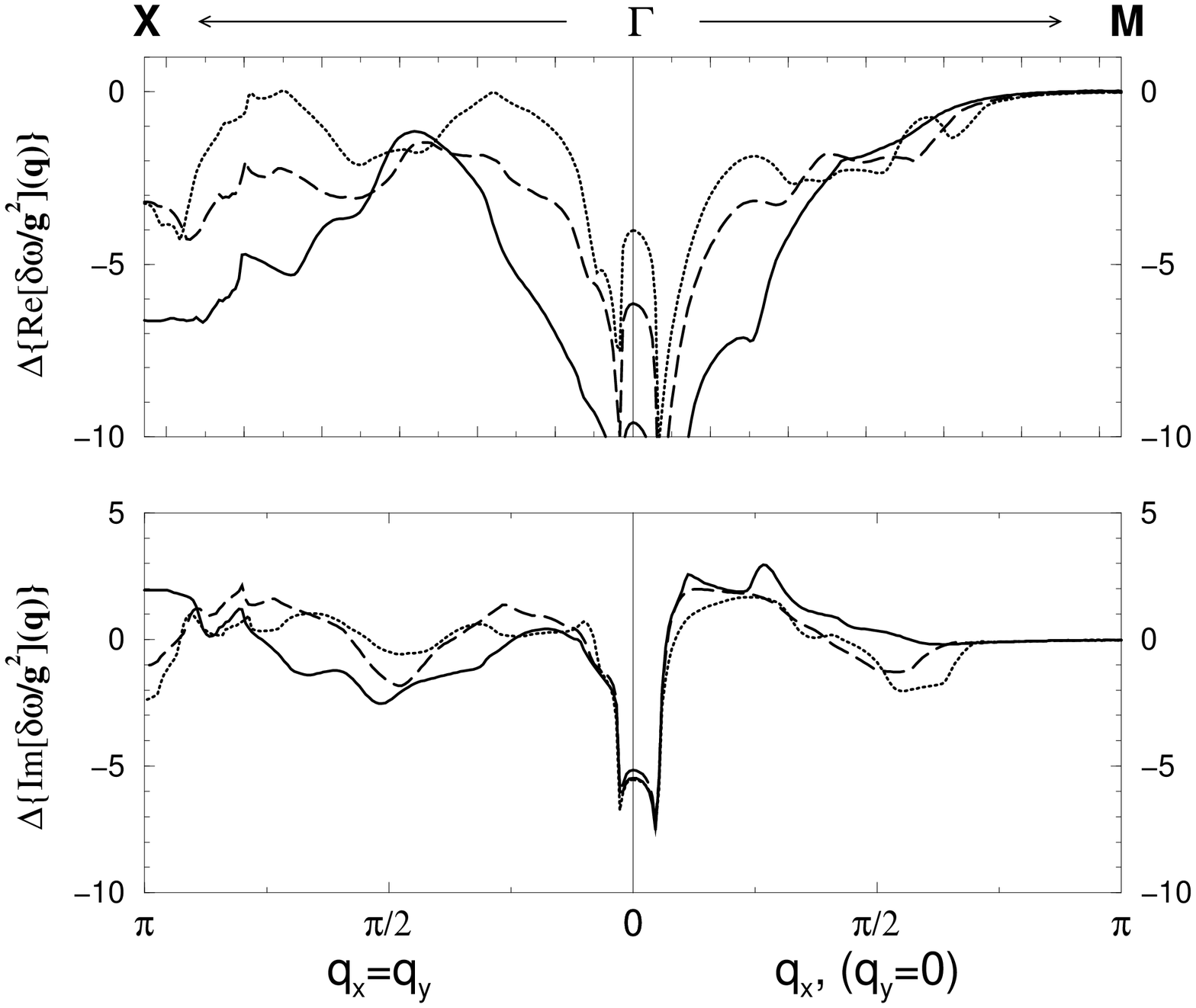,width=7.5cm}}}
\end{center}
\vspace{-0.2cm}
{\small FIG. 4. SC induced change in the q-dependent frequency shift
and linewidth broadening for 
 $\omega_{a} = 2\Delta(0)- \Delta(0)/2$.  Labeling of the curves is the 
same than in Fig.1}
\end{figure}

they are only observable when the condition $q_z=0$ is fulfilled.
However, since the latter is experimentally hard
to realize we disregard these features in our following analysis.

Consider first the $(0,0) \to (\pi,0)$ ($\Gamma \to M$) 
scan where the pseudogap induced 
change in $Re \Pi(q,\omega)$ is most significant for wave vectors
up to $q \sim (\pi/2,0)$. In fact, for zero DDW gap the system is rather
susceptible for perturbations within this wave vector range since 
the corresponding scattering acts in the high-density $(\pi,0)$ region
of the underlying dispersion. However, it is exactly this part of the 
Brillouin zone where the DDW state induces a gap which in turn leads 
to the decrease in the charge susceptiblity for $q_x < \pi/2$.
As a consequence the most pronounced change in the SC induced phonon
frequency shift now occurs at around $q \approx (\pi/2,0)$ 
which reflects itself in an increase of the linewidth around the 
same wave vector.
For the $(0,0)\to (\pi,\pi)$ ($\Gamma \to X$) scan the same
arguments than above can be applied in order to understand
the suppression of small q frequency shifts upon increasing the
DDW gap. Moreover, along the diagonals also the large wave vector
phonon renormalization is affected. Thus the opening of the pseudogap 
strongly reduces the SC induced frequency change over large parts
of the $\Gamma \to X$ direction.

Therefore, our results can account for some of the anomalies of the
340cm$^{-1}$ phonon observed by inelastic neutron scattering 
\cite{PYKA,REZNIK}. Most interestingly it was found that at low temperature 
the SC induced change in phonon frequency is largest along the $(1,0)$ 
direction for wave vectors up to $q \sim (\pi/2,0)$. Then the softening
progressively vanishes towards the zone boundary and as a consequence the
linewidth aquires a maximum at $q \approx  (\pi/2,0)$.
On the other hand much smaller softening and no linewidth broadening
has been detected along the diagonal direction.
Thus from our analysis one may deduce that the measured linewidth broadening
along the $(1,0)$ direction is at least partially due to the 
presence of the pseudogap.
This feature together with the reduced frequency renormalization 
of phononic Raman modes upon underdoping supports the point of view,
that the pseudogap also persists below T$_c$ and may not (only) originate
from pair fluctuations.

Obviously for a quantitative comparison a more detailed
analysis for the q-dependent electron-phonon coupling is necessary 
which for the 340cm$^{-1}$ mode has been done in Ref. \cite{DEVER}.
In addition, we have restricted ourselves to the purely two-dimensional
BCS case. However, at least the qualitative features of
our investigations should be robust with respect to these approximations
and may provide a basis for a more quantitative analysis of phonon
renormalizations under the presence of an anisotropic  pseudogap.

The authors would like to thank A. Bill for helpful discussions and
a critical reading of the manuscript.


\begin{references}
\bibitem{NOTE} Note that for zero pseudogap
the double peak structure in the linewidth is due to the small separation
of van Hove singularity and SC 'leading gap' edge.
\end{references}
\end{document}